\documentclass[12pt]{article}
\usepackage{epsfig,amsfonts,amssymb}
\usepackage{hyperref}
\pdfoutput=1

\usepackage{cite}
\usepackage{cite}

\usepackage{comment}

\def\ZZZ{{\hbox{ Z\kern-1.6mm Z}}}
\def\RRR{{\hbox{ R\kern-2.4mm R}}}
\def\CCC{{\hbox{ C\kern-2.0mm C}}}
\def\zzz{{\hbox{z\kern-1mm z}}}

\newcommand{\qeq}{{\hbox{=\kern-2.3mm ? \kern.5mm }}}
\renewcommand{\qeq}{=}

\newcommand{\OO}{{\cal O}}

\newcommand{\wt}{\widetilde}
\newcommand{\wh}{\widehat}

\newcommand{\NN}{{\cal N}}
\newcommand{\TT}{{\cal T}}

\newcommand{\be}{\begin{equation}}
\newcommand{\ee}{\end{equation}}
\newcommand{\ben}{\begin{eqnarray}\displaystyle}
\newcommand{\een}{\end{eqnarray}}

\newcommand{\refb}[1]{(\ref{#1})}
\newcommand{\p}{\partial}
\newcommand{\sectiono}[1]{\section{#1}\setcounter{equation}{0}}

\def\one{{\hbox{ 1\kern-.8mm l}}}
\def\zero{{\hbox{ 0\kern-1.5mm 0}}}

\newcommand{\bea}[1]{\begin{eqnarray}\label{#1} }
\newcommand{\eea}{\end{eqnarray}}

\newcommand{\eqref}{\refb}

\usepackage{bm}
\usepackage[table]{xcolor}

\begin{document}

\baselineskip 24pt

\begin{center}

{\Large \bf D-instanton Induced Effective Action and its Gauge Invariance}

\end{center}

\vskip .6cm
\medskip

\vspace*{4.0ex}

\baselineskip=18pt

\centerline{\large \rm Ashoke Sen}

\vspace*{4.0ex}

\centerline{\large \it International Centre for Theoretical Sciences - TIFR 
}
\centerline{\large \it  Bengaluru - 560089, India}


\vspace*{1.0ex}
\centerline{\small E-mail:  ashoke.sen@icts.res.in}

\vspace*{5.0ex}

\centerline{\bf Abstract} \bigskip

The effect of D-instantons on closed string scattering amplitudes may be encoded into an 
effective action obtained by integrating out the (transient) open string fields. In order that
this respects the gauge invariance of the theory, the sum of the 
perturbative closed string field theory
action and the D-instanton induced effective action must satisfy the quantum
Batalin-Vilkovisky master equation. In a previous paper this was proved for the effective action
induced by up to two instantons. We generalize the proof for arbitrary number of D-instantons.

\vfill \eject

\tableofcontents

\sectiono{Introduction} \label{s0}

The action of closed string field theory satisfies the quantum Batalin-Vilkovisky (BV)
master equation\cite{bv1,bv}
encoding the gauge invariance of the theory\cite{Zwiebach:1992ie,Sen:2024nfd}. 
However besides the perturbative 
contribution to the action, closed string field theory also receives non-perturbative
contributions to the action from D-instantons. 
A systematic algorithm for computing D-instanton contribution
to the closed string field theory action was described in \cite{Sen:2020ruy}, but the explicit 
construction of the action was carried out only up to two instanton contribution.
It was also checked that the action, so constructed, satisfies the BV master equation
up to corrections involving three instanton terms.

Our goal in this paper will be to construct the action of closed string field theory after
taking into account the effect of arbitrary number of instantons. We also show that the
action so constructed satisfies the quantum BV master equation to all orders in the
instanton expansion. Our main result, describing the instanton induced effective action
due to arbitrary number of identical D-instantons, is given in \refb{e21}.
Its generalization to arbitrary numbers of non-identical D-instantons, is given in
\refb{e23alternew}.

One immediate application of our results is the proof of gauge invariance of the closed
string amplitudes even after the inclusion of D-instanton effects.
Since for computing D-instanton contribution to
closed string amplitudes
we need to modify the world-sheet rules, separating out the zero mode integrals and
doing these integrals at the end, the usual perturbative proof of gauge invariance
showing the decoupling of BRST trivial states no longer holds. However the fact that the
closed string effective action satisfies the quantum BV master equation immediately
implies the gauge invariance of on-shell closed string amplitudes.

\sectiono{Review and the statement of the problem} \label{s1}

In this section we shall review the relevant results of \cite{Sen:2020ruy} using a
somewhat different notation and paying special attention to the zero mode
integrals.

Let us consider closed strings interacting with some fixed number
of  D-branes. 
Even though our eventual goal is to study D-instantons, let us first consider the case
where the D-branes extend along more than two non-compact space-time directions so
that they can be regarded as fixed background. 
The observables in this system
are scattering amplitudes involving external open and closed strings and these
amplitudes can be reproduced by a gauge invariant field theory of open and closed
strings\cite{Zwiebach:1997fe}. 
The interaction vertices of this field theory are expressed as integrals over
appropriate subspaces of the 
moduli spaces of punctured, {\it connected} 
Riemann surfaces, with the integrand given in terms of
correlation
functions of off-shell open and closed strings inserted at the
punctures, together with some additional insertions of ghost and / or picture 
changing operators. The subspaces of the moduli space on which we carry out the
integration avoid all degenerate Riemann surfaces and hence the integrals 
do not suffer from
any divergences.  More details have been reviewed in \cite{Sen:2024nfd}.

There is a standard procedure, also reviewed in \cite{Sen:2024nfd}, 
by which we can integrate out a set
of degrees of freedom of the theory and write down an expression for the effective action
involving the rest of the degrees 
of freedom\cite{Sen:2016qap,Kajiura:2001ng,Kajiura:2003ax}. 
In particular, if we are interested in  amplitudes
involving only the external closed strings, we can integrate out the open strings and write
down an effective action for closed string fields only. The interaction vertices of the
resulting effective closed string field theory are still given by integrals over
appropriate subspaces of the 
moduli spaces of punctured connected
Riemann surfaces. The difference is that now the integrand 
has correlation functions of external closed strings only and the integration region
now includes degenerate Riemann surfaces involving open string degeneration but
still avoids degenerate Riemann surfaces involving closed string degeneration.
Since the integrals may get divergent contributions from the region of the moduli space
where there is open string degeneration,
the definition of the effective action 
given above is somewhat formal. 
For D-instantons that will be of interest to us, the solution to this problem has been
well understood and reviewed in \cite{Sen:2024nfd}. 
We shall discuss some consequences of this when we turn to D-instantons.

Let $S_{(r)}$ denote the total closed string field theory effective action for $r$ identical 
D-branes.\footnote{Here identical means D-branes that differ from each
other at most by their open string moduli {\it e.g.} the position of the D-brane in
transverse directions. They have the same tension.} $S_{(r)}$ includes the contribution from {\it connected}
Riemann surfaces without boundaries as well as {\it connected} Riemann surfaces with
boundaries lying on any of the $r$ D-branes.
$S_{(0)}$ will denote the closed string field theory action without any D-brane in the
background.
Then $S_{(r)}$ will satisfy the quantum BV
master equation\cite{Sen:2024nfd}
\be\label{e1}
{1\over 2} \, \{S_{(r)}, S_{(r)}\} + \Delta S_{(r)} = 0\, , \qquad \hbox{for $r\ge 0$}\, ,
\ee
where for a pair of functions $F$ and $G$ of closed string fields $\{\psi^\alpha\}$ and their
anti-fields $\{\psi_\alpha^*\}$, we have
\be\label{e2}
\{F, G\} = {\p_R F \over \p \psi^\alpha} \, {\p_L G\over \p \psi_\alpha^*} -
{\p_R F \over \p \psi_\alpha^*} \, {\p_L G\over \p \psi^\alpha}, \qquad
\Delta F = (-1)^{\psi_\alpha} {\p_L \over \p \psi^\alpha} \, 
\left({\p_L F\over \p \psi_\alpha^*}\right)\, ,
\ee
where $\p_L$ and $\p_R$ denotes left and right derivatives respectively and 
$(-1)^{\psi_\alpha}$ takes value 1 ($-1$) for Grassmann even (odd) $\psi_\alpha$. 
For even functions $F$, we have the useful property,
\be\label{ed}
\Delta e^F = {1\over 2} \{F,F\} \, e^F + \Delta F \, e^F\,  .
\ee
Using this, \refb{e1} may be written as
\be\label{e1alt}
\Delta \, e^{S_{(r)}} = 0\, .
\ee

We shall see later that for D-instantons the definition of $S_{(r)}$ given above
requires some modification, so it will be useful to give a formal definition of $S_{(r)}$
that is valid in all cases. Let 
$\wt S^{(r)}$ denote the full BV master action of open-closed string field theory in the
presence of $r$ D-branes
before we integrate out any fields. Then $e^{S_{(r)}}$ is defined as the
integral of $e^{\wt S_{(r)}}$ over half of the open string fields, the other half
being fixed by `gauge choice'.  $e^{\wt S_{(r)}}$ satisfies the BV master equation
$(\Delta+\Delta_{open})e^{\wt S_{(r)}}=0$\cite{Zwiebach:1997fe,Sen:2024nfd}. 
Integrating this over half of the open
string fields, we see that the term involving $\Delta_{open}e^{\wt S_{(r)}}$ integrates
to zero since it is a total derivative, and we are led to \refb{e1alt}. \refb{e1} then
follows as a consequence of \refb{ed} after defining $S_{(r)}$ as $\ln e^{S_{(r)}}$.
If we construct $S_{(r)}$ using perturbation theory, then by the standard rules
$S_{(r)}$ will be 
given by the sum of connected Feynman 
diagrams, with the fields that are integrated out appearing in the propagators.
In string theory in the presence of D-branes,
the sum over connected Feynman diagrams
translates to sum over connected world-sheets with boundaries, leading to the
description of $S_{(r)}$ given earlier.

We shall now introduce some new definitions.
We shall denote by $\wh S_i$ the contribution to the closed string effective action from
the sum of all
{\it connected} world-sheets with all boundaries on the $i$-th D-brane.
$\wh S_{ij}$ will denote the contribution from
the sum of all
connected world-sheets with each boundary on either the $i$-th D-brane or the
$j$-th D-brane, at least one boundary on the $i$-th D-brane and at
least one boundary on the $j$-th D-brane. Similar definitions may be given for
$\wh S_{i_1\cdots i_k}$ when $k$ or larger 
number of D-branes are present. 
It follows from the definition that $\wh S_{i_1\cdots i_k}$ is symmetric under the exchange
of the indices.
Then we have, for example,
\be\label{e25new}
S_{(1)}= S_{(0)} + \wh S_1\, , \qquad S_{(2)} = S_{(0)} + \wh S_1 + \wh S_2 +
\wh S_{12}, \qquad S_{(3)} = S_{(0)}+ \wh S_1 + \wh S_2 + \wh S_3+
\wh S_{12} + \wh S_{23} +\wh S_{13} +\wh S_{123}\, ,
\ee
etc.

Let us now turn to the
case of D-instantons. 
As usual, we shall use a general definition of D-instantons -- D-branes that satisfy
Dirichlet boundary condition for all {\it non-compact} space-time directions.
In this case the construction of the closed string effective action requires some
modification.
There are several reasons for this that we
enumerate below. 
\begin{enumerate}
\item
Unlike in the case of extended D-branes where the contribution to the closed string
effective action comes from a sum over connected world-sheets with boundaries, 
for computing D-instanton contribution to the closed string effective action,
we have 
to include contribution from {\it disconnected} world-sheets\cite{Polchinski:1994fq}.
This is related to the infrared divergences associated with open string degeneration
that was alluded to earlier. The problem comes from integration over the open string zero
modes, including the D-instanton position. 
The presence of these zero modes makes the building blocks
$\wh S_{i_1\cdots i_k}$ ill defined.
The remedy is to remove the contribution of these zero modes from the internal
open string propagators (and include a contribution from an out of Siegel gauge
mode propagator\cite{Sen:2020eck}), 
leading to modified rules for world-sheet computation.
This gives 
finite $\wh S_{i_1\cdots i_k}$. We shall denote this by $\wh S'_{i_1\cdots i_k}$ to
indicate that the rules for world-sheet computation have been modified.
However since the construction of the closed string effective action requires
integrating out all the open string modes,
the zero mode integrals must be performed
at the end. 
As a result, disconnected world-sheets that end on a common D-instanton 
will have their zero mode integrals performed together and will contribute
to the closed string effective action. We shall call them {\it target space connected
contributions}. 
Since the overall energy momentum conservation is implemented
only after integration over the zero modes, a pair of disconnected world-sheets that
are target space connected will have one overall energy-momentum conserving
delta function instead of separate energy-momentum conservation associated with
individual world-sheets.
For example the product of $\wh S'_{12}$ and $\wh S'_{23}$ 
is target space
connected since both end on the D-instanton 2 at one of their boundaries.
Hence its contribution is included in the closed string effective action $S$ after
performing integration over the zero modes of instantons 1 and 2.
But not all disconnected world-sheets are allowed, {\it e.g.} the product of
$\wh S'_{12}$ and $\wh S'_{34}$ is not
target space connected and does not contribute to the four instanton
effective action.\footnote{Instead, it should be viewed as a contribution to 
the square of the two instanton
effective action.}
However if we multiply this by $\wh S'_{13}$ then the world-sheets
are connected in the target space and
this product becomes an allowed four instanton contribution to the effective
action after we integrate over the zero modes of the instantons 1,2,3 and 4. 
\item Since instantons have finite action, they do not represent fixed backgrounds but
correspond to
saddle points of the path integral that contribute to the closed string amplitudes in the
vacuum. Hence 
we must sum over the contributions from different number of instantons.
An $r$ instanton contribution to the effective action 
is multiplied by a factor of $\NN^r$ where
$\NN=e^{-\TT}$, $\TT$ being the D-instanton tension.\footnote{In 
\cite{Sen:2016qap,Sen:2024nfd} we included
also the exponential of the annulus amplitude in the definition of $\NN$ but here we
shall include this in the action.} 
\item Since  the path integral over the open string fields on the D-instanton includes
integration over the D-instanton positions and other moduli, after the integration different
D-instantons become indistinguishable. This gives rise to an extra symmetry factor
$1/r!$ for an $r$-instanton contribution.
\end{enumerate}

With these rules the contribution to the closed
string effective action up to two instanton sector
can be written as\cite{Sen:2016qap,Sen:2024nfd}
\be \label{etwofirst}
S\equiv S_{(0)} + \NN \int_z \left(e^{\wh S'_1}-1\right) 
+ {1\over 2} \, 
\NN^2 \, \int_z\, 
\left(e^{\wh S'_{12} +\wh S'_1 + \wh S'_2} - 1 - e^{\wh S'_1+\wh S'_2}  + 1 \right) + \OO(\NN^3)  \, ,
\ee
where $\int_z$ signs on the right hand side of \refb{etwofirst} denote 
integration over
the zero modes of the instantons. In order to determine what zero modes we
should integrate over, we should expand the exponentials in power series
in $\wh S'_{i_1\cdots i_k}$ and in each term carry out the integration over the
zero modes of the instantons that appear in the product. For example for $\wh S'_{12}$
we integrate over the zero modes of instantons 1 and 2 while for $\wh S'_{12}
\wh S'_{23}$ we integrate over the zero modes of the instantons 1, 2 and 3.
In \refb{etwofirst} $S_{(0)}$ is the perturbative contribution to the effective action and the
term proportional to $\NN^r$ represents the $r$ instanton contribution to the
effective action. $e^{\wh S'_1}-1$, expanded in a power series in
$\wh S'_1$, generates connected and 
disconnected world-sheets where each connected
component has at least one boundary and all the boundaries end on a single
D-instanton. 
All of these diagrams are connected in the target space and represent 
one instanton
contribution to the effective action. This explains why they are 
multiplied by $\NN$.
To see how the coefficient of $\NN^2$ arises, 
we note that the $e^{\wh S'_{12} +\wh S'_1 + \wh S'_2} - 1$  
factor, expanded in a power series in
$\wh S'_{12}$, $\wh S'_1$ and $\wh S'_2$, 
generates connected and disconnected world-sheets, where each connected
component has at least one boundary lying on a
D-instanton. 
As long as we have terms with at least one factor of $\wh S'_{12}$, the diagram is
connected in the target space and should be included in the closed string effective
action. However the terms without any factor of $\wh S'_{12}$ are either disconnected
in the target space or has all its boundaries lying on one instanton. These contributions
are not genuine two instanton contribution to the effective action and must be
removed. The subtraction term $e^{\wh S'_1+\wh S'_2}  - 1$ precisely achieves
this.

We shall now express \refb{etwofirst} in terms of the $S_{(r)}$'s  whose general definition
was given below \refb{e1alt}.
For D-instantons we cannot use \refb{e25new} since the contribution from the
open string zero modes are removed from the definition of $\wh S'_{i_1\cdots i_k}$.
Instead,
we can express $S_{(r)}$ via,
\be\label{e25newer}
e^{S_{(1)}}= \int_z e^{S_{(0)}} e^{\wh S'_1}\, , \qquad e^{S_{(2)} }= \int_z e^{S_{(0)}} 
e^{\wh S'_1 + \wh S'_2 +
\wh S'_{12}}, \qquad
e^{S_{(3)}} = \int_z e^{S_{(0)}} e^{\wh S'_1 + \wh S'_2 + \wh S'_3+
\wh S'_{12} + \wh S'_{23} +\wh S'_{13} +\wh S'_{123}}
\ee
etc.
Due to the argument given below \refb{e1alt}, $S_{(r)}$
satisfies the BV master equation
\refb{e1alt}, \refb{e1}.  Using \refb{e25newer}, 
we can express \refb{etwofirst}
as
\be\label{e5}
S = S_{(0)} + \NN \left(e^{S_{(1)}-S_{(0)}}-1\right) 
+ {1\over 2} \, 
\NN^2 \, \left(e^{S_{(2)} - S_{(0)}}  - e^{2(S_{(1)}-S_{(0)})} 
 \right) + \OO(\NN^3)  \, ,
\ee
where we used the fact that
after carrying out the zero mode integrals,
$e^{\wh S'_1}$ and $e^{\wh S'_2}$ become identical to $e^{S_{(1)}-S_{(0)}}$.

For consistency, $S$ given in
\refb{e5} should satisfy the BV master equation. To prove this, we
use \refb{e1} with $r=0,1,2$ and  \refb{ed}
to get\cite{Sen:2016qap,Sen:2024nfd}
\be\label{e15}
{1\over 2} \, \{S, S\} + \Delta \, S = \OO(\NN^3)\, ,
\ee
as required.

We are now ready to state the problem. This can be divided into two parts:
\begin{enumerate}
\item We need to find the $r$ instanton contribution to $S$, i.e.\ find 
the coefficient of the $\NN^r$ term
in \refb{e5}. This requires us to generalize \refb{etwofirst} or \refb{e5} for larger
number of instantons. The subtraction terms now take more complicated
form, {\it e.g.} for three instantons we need to keep terms that contain at least
one factor of $\wh S'_{123}$, but we need to also keep terms that contain $\wh S'_{12}
\wh S'_{23}$ or $\wh S'_{12}
\wh S'_{13}$ or $\wh S'_{13} \wh S'_{23}$ as a factor.
\item Once we have found $S$,
we need to prove that the left hand side of \refb{e15} vanishes exactly.
\end{enumerate}

\sectiono{Multi-instanton contribution to the effective action and its
gauge invariance} \label{s2}

In this section we shall answer the questions posed at the end of the last section. We
claim that the full action takes the form
\be\label{e21}
S = S_{(0)} + \ln \left( 1+ \sum_{r=1}^\infty {1\over r!} \, \NN^r \, e^{S_{(r)} - S_{(0)}}\right)
- \NN\, .
\ee
The explicit computation of $e^{S_{(r)}}$
will involve sum of all connected and disconnected world-sheets in the presence of
$r$ D-instantons with the modified world-sheet rules, 
followed by integration over the zero modes of these instantons. 

Before giving a proof of \refb{e21},
let us first verify that its expansion to order $\NN^2$ agrees with
\refb{e5}. By expanding \refb{e21} to order $\NN^2$, we get
\be \label{e32a}
S = S_{(0)} + \NN (e^{S_{(1)} - S_{(0)}}-1) + {1\over 2!} \, \NN^2 \, \left(e^{S_{(2)} - S_{(0)}}
- e^{2(S_{(1)} - S_{(0)})} \right)+\OO(\NN^3) \, ,
\ee
in agreement with \refb{e5}.

Let us now turn to the proof of \refb{e21}. For this we rewrite this equation as
\be\label{e23}
e^S = e^{-\NN} e^{S_{(0)}} \left[1 + \sum_{r=1}^\infty {1\over r!}\, \NN^r\,  e^{S_{(r)}
- S_{(0)}}
\right]\, .
\ee
On the left hand side, since $S$ is supposed to be
the effective action, the expansion of $e^S$ should generate
all possible combination of interaction vertices that arise in a Feynman diagram -- with
the understanding that the actual Feynman diagrams are constructed by joining the
vertices to each other and to the external lines by propagators. 
Therefore this should
correspond to all possible combinations of connected and disconnected world-sheets
with external closed strings.\footnote{Here by `all possible' world-sheets we mean only
those Riemann surfaces
that appear in the construction of the interaction vertices of closed string effective
field theory. Other Riemann surfaces 
that are constructed from these by sewing with closed
string propagators are not included.}
Thus checking \refb{e23} amounts to checking that the right hand side generates all
possible world-sheets, including disconnected world-sheets that are not even connected
in the target space. First consider the case of world-sheets without boundaries.
These come from terms that contain only factors of $S_{(0)}$ but no factors of 
$S_{(r)}-S_{(0)}$. Picking the 1 term in the expansion of $e^{(S_{(r)}
- S_{(0)})}$ in \refb{e23}, we see that the coefficient of the $e^{S_{(0)}}$ term is
\be
e^{-\NN} \left[ 1 + \sum_{r=1}^\infty {1\over r!}\, \NN^r\right] = 1\, .
\ee
Therefore the right hand side of \refb{e23} has the contribution of connected and
disconnected world-sheets, each without any boundary, with the correct coefficient.

Next we focus on the $s$ instanton contribution from the right hand side of \refb{e23}.
Here by $s$-instanton contribution we mean the contribution from
those connected or disconnected
Riemann surfaces for which each of the boundaries lies on one of the $s$ instantons
and each of the $s$ instantons has at least one boundary lying on it.
One source of such terms is the expansion of $e^{S_{(s)}
- S_{(0)}}-1$. However the expansion of 
$e^{S_{(r)}
- S_{(0)}}-1$ for any $r\ge s$ also contains $s$ instanton contribution, since in the
product of the $S_{(r)}-S_{(0)}$ factors, we 
have a contribution where all the boundaries lie on $s$ of the $r$ instantons and each
of the $s$ instantons has at least one boundary lying on it.
Since $s$ instantons can be chosen from
$r$ instantons in ${r\choose s}$ ways, the net coefficient of the $s$ instanton contribution
to the amplitude will be given by
\be
e^{-\NN} \sum_{r\ge s} {1\over r!} \, \NN^r \, {r\choose s} =
e^{-\NN} \sum_{r\ge s} {1\over s! (r-s)!} \NN^r = {1\over s!} \NN^s\, .
\ee
This is precisely the desired contribution, since we need a factor of $\NN^s$ to
accompany the $s$ instanton contribution and the $1/s!$ term reflects the fact
that we have $s$ identical instantons. 

This establishes that the right hand side of \refb{e23} gives the perturbative and
instanton contribution to $e^S$ with the correct coefficient. 

It is instructive to study the coefficient of the $\NN^3/3!$ term in \refb{e21}. It is given by
\be\label{encube}
\left[ e^{S_{(3)}-S_{(0)}} - 3\, e^{S_{(2)}-S_{(0)} + S_{(1)}-S_{(0)}} 
+ 2\, e^{3 (S_{(1)}-S_{(0)})}\right]\, .
\ee
With the definition of $\wh S'_{i_1\cdots i_k}$ given in section \ref{s1}, 
\refb{encube} may be
expressed as:
\be
\int_z \left[e^{\wh S'_{123}+\wh S'_{12}+\wh S'_{23}+\wh S'_{13} +\wh S'_1+\wh S'_2+\wh S'_3} 
-e^{\wh S'_{12}+\wh S'_1+\wh S'_2 + \wh S'_3}
- e^{\wh S'_{13}+\wh S'_1+\wh S'_3 + \wh S'_2} 
- e^{\wh S'_{23}+\wh S'_2+\wh S'_3 + \wh S'_1} + 2\, 
 e^{\wh S'_1+\wh S'_2 + \wh S'_3}\right]\, .
\ee
Expanding this in a power series in $\wh S'_i$, $\wh S'_{ij}$ for $1\le i,j\le 3$
and $\wh S'_{123}$, we can verify that each term in the expansion has one of the
following four factors: $\wh S'_{123}$, $\wh S'_{12}\wh S'_{23}$,
$\wh S'_{12}\wh S'_{13}$ or $\wh S'_{13}\wh S'_{23}$. This shows that each
world-sheet diagram contributing to \refb{encube} is connected in the target space
as desired.

We now turn to the second part of the problem, namely to show
that $S$ defined in \refb{e21}
satisfies the BV master equation
\be\label{ebvfin}
{1\over 2} \{S,S\} + \Delta S=0\, .
\ee
For this we express \refb{e23} as
\be\label{e23alt}
e^S = e^{-\NN}  \left[e^{S_{(0)}} + \sum_{r=1}^\infty {1\over r!}\, \NN^r\,  e^{S_{(r)}
}
\right]\, .
\ee
It now follows from \refb{e1alt} that we have
\be
\Delta e^S =0\, .
\ee
This leads to \refb{ebvfin} as a consequence of \refb{ed}.

\sectiono{Non-identical D-instantons}

The generalization of these results to non-identical D-instantons is straightforward.
Let us suppose that the theory contains $k$ different kinds of instantons of tensions
$\TT_1,\cdots,\TT_k$ and let $\NN_i=e^{-\TT_i}$ for $1\le i\le k$.
Let us denote by $e^{S_{(r_1,r_2,\cdots r_k)}}$ the sum over
contributions from all connected and
disconnected world-sheets in the presence of $r_1$ instantons of the first
kind,  $r_2$ instantons of the second kind etc., followed by integration over the
D-instanton positions and other zero modes. 
$S_{(r_1,r_2,\cdots r_k)}$ 
is then defined to be the logarithm of $e^{S_{(r_1,r_2,\cdots r_k)}}$.
Then
$S_{(r_1,r_2,\cdots r_k)}$ satisfies the BV master equation
\be\label{e49}
{1\over 2} \{S_{(r_1,r_2,\cdots r_k)}, S_{(r_1,r_2,\cdots r_k)}\} +
\Delta S_{(r_1,r_2,\cdots r_k)}=0 \qquad \Leftrightarrow \qquad
\Delta \, e^{S_{(r_1,r_2,\cdots r_k)}} = 0\, .
\ee

Our goal is to construct the effective action for closed strings after taking into account
the effect of the weight factors $\prod_{i=1}^{k} \NN_i^{r_i}$ and including the 
contribution from the disconnected world-sheets that are target space connected. 
By following the same logic
that established the validity of \refb{e23}, we arrive at 
\be\label{e23alter}
e^S = e^{-\NN_1-\NN_2 -\cdots -\NN_k}   \sum_{r_1,\cdots , r_k=0}^\infty {1\over r_1! r_2!\cdots r_k!}\, \NN_1^{r_1} \NN_2^{r_2} \cdots \NN_k^{r_k}\,  
e^{S_{(r_1,r_2,\cdots r_k)}
}
\, .
\ee
This can also be written as
\be\label{e23alternew}
S =  S_{(\vec 0)} + \ln \left[1 +  \sum_{r_1,\cdots , r_k=0\atop
(r_1,\cdots , r_k)\ne \vec 0}^\infty {1\over r_1! r_2!\cdots r_k!}\, \NN_1^{r_1} \NN_2^{r_2} \cdots \NN_k^{r_k}\,  
e^{S_{(r_1,r_2,\cdots r_k)} - S_{(\vec 0)}}\right] -\NN_1-\NN_2 -\cdots -\NN_k 
\, .
\ee
Here $S_{(\vec 0)}$ denotes the perturbative contribution to the effective action.

We now see that as a consequence of \refb{e49} and \refb{e23alter}, we have
\be
\Delta \, e^S = 0 \qquad \Leftrightarrow \qquad {1\over 2} \{S,S\} + \Delta S=0\, .
\ee
This establishes that the full closed string effective action induced by multiple D-instantons 
of different kinds satisfy the BV master equation.

\bigskip

\noindent{\bf Acknowledgement:} 
We thank Barton Zwiebach for useful discussions and comments and suggestions
on an earlier draft
of the manuscript that helped improve the presentation.
This work was supported by the ICTS-Infosys Madhava 
Chair Professorship, the J. C. Bose fellowship of the Department of Science
and Technology, India
and the Department of Atomic Energy, Government of India, under project no. RTI4001.

\end{document}